\documentclass[final,3p,times,twocolumn]{elsarticle}

\usepackage{float}
\usepackage{amsfonts}
\usepackage{amsmath}    
\usepackage{amssymb}
\usepackage{graphicx}   
\usepackage{xcolor}      
\usepackage{subfigure}  
\usepackage{braket}
\journal{Physics Letters A}
\begin{document}
\begin{frontmatter}
\title{New Bell inequalities  for three-qubit pure states}

\author{Arpan Das}
\ead{arpandas@iopb.res.in}
\author{Chandan Datta}
\ead{chandan@iopb.res.in}
\author{Pankaj Agrawal}
\ead{agrawal@iopb.res.in}
\address{Institute of Physics - Sachivalaya Marg, Sainik School, Bhubaneswar 751005, Odisha, India. }
\address{Homi Bhabha National Institute - Training School Complex, Anushakti Nagar, Mumbai 400085, India.}

\begin{abstract}
We introduce a set of Bell inequalities for a three-qubit system. Each inequality within this set is violated by all generalized GHZ states. More entangled a generalized GHZ state is,  more will be the violation. This establishes a relation
between nonlocality and entanglement for this class of states.
Certain inequalities  within this set are  violated by pure biseparable states. We also provide numerical evidence that at least one of these Bell inequalities
is violated by a pure genuinely entangled state. These Bell inequalities can distinguish between separable,
biseparable and genuinely entangled pure three-qubit states. 
We also generalize this set to $n$-qubit systems and may be suitable to characterize the entanglement 
of $n$-qubit pure states.
\end{abstract}
\end{frontmatter}
\section{Introduction}

Recent loophole free tests \cite{free} of Bell inequalities have firmly established that certain correlations of 
quantum states cannot be explained by a local hidden variable (LHV) model. Bell \cite{bell1964} had demonstrated
that correlations in a local-realistic theory must obey an inequality -- Bell inequality. He considered 
a simple situation of two spin-$\frac{1}{2}$ particles in a singlet state and showed that quantum mechanics is not compatible with local-realistic theories. Since then, this result has been generalized
in many different directions \cite{brunnerRMP}.  In the simplest situation (two parties, two measurement settings and two
outcomes per setting), all facet Bell inequalities are equivalent to the Clauser-Horner-Shimony-Holt (CHSH) 
inequality \cite{chsh}. This is because CHSH inequality is the only nontrivial facet of Bell polytope for such a scenario \cite{brunnerRMP}. According to Gisin's theorem \cite{gisin}, all pure bipartite entangled states violate this inequality.
The maximum allowed expectation value of CHSH operator for a quantum system is $2 \sqrt{2}$. This
bound is known as the Tsirelson's bound \cite{tsirelson}. However, the  violation of a Bell inequality
is only sufficient criteria for certifying entanglement but not a necessary one. Even in the case of two-qubit
states, there is a phenomena of hidden nonlocality, which is typified by Werner state \cite{werner}.
However, this phenomena occurs for mixed states. In this paper, we will consider pure multipartite states. Unlike a pure bipartite state, the relationship
between Bell nonlocality and entanglement is far from simple \cite{coll}. 
For three-qubit states, we will adopt following terminology.
 A state $\ket{\psi}$ is a pure separable or product state if it can be written in the 
form $\ket{\psi_1}\otimes \ket{\psi_2}\otimes\ket{\psi_3}$, a pure biseparable state if it can be 
written as $\ket{\psi_1}\otimes \ket{\psi_{23}}$ or in other permutations and is genuinely 
entangled if it cannot be written in a product form. Above classification is based on types of entanglement present in the state. Idea of non-separability according to Bell locality comes from the inability 
of construction of a LHV model for observed correlations.
 In this case, the quantum mechanical description is not presupposed. For a system of three particles, if the joint probability can be written as,
\begin{equation*}
P(a_1, a_2, a_3)=\int d\lambda\rho(\lambda)P_1(a_1|\lambda)P_2(a_2|\lambda)P_3(a_3|\lambda),
\end{equation*}
where $P_i(a_i|\lambda) (i=1,2,3)$ is the probability of yielding the result $a_i$, when a measurement $A_i$ is 
done on the particle with the local hidden variable $\lambda$, then the model is the well known LHV model.  The intermediate case is the hybrid local-nonlocal 
model,  first considered by Svetlichny \cite{svet}, where there is an arbitrary 
nonlocal correlation between two of the three particles but only local correlations between these two 
and the third particle. The last situation is genuine tripartite nonlocality, where three 
particles are allowed to share arbitrary correlations. Whereas
Svetlichny's inequality allowed arbitrary nonlocal correlation between two parties, a more refined and 
strictly weaker definition of genuine tripartite nonlocality was introduced in \cite{bancal}. By analyzing 
the no-signalling polytope,  they found a set of $185$ inequivalent facet inequalities and numerically conjectured 
that every genuine tripartite entangled states show violations within this set and hence 
they are also genuinely tripartite nonlocal according to their definition.\\

In the case of three qubits, violation of Mermin-Ardehali-Belinskii-Klyshko (MABK) inequalities \cite{mermin,bec} gives sufficient criteria to distinguish separable states from entangled ones. But it is not a necessary condition as there are states, which do no violate MABK inequalities but have genuine tripartite entanglement \cite{gen}. {\'S}liwa \cite{sli} constructed the Bell polytope i.e all tight Bell inequalities for three parties and two dichotomic measuements per party, where Mermin inequality is one of the facets. More precisely, in \cite{gen} it was shown that the $n$-qubit state, $\ket{\psi}=\cos\alpha \ket{0...0}+\sin\alpha \ket{1...1}$ (we will call it generalized GHZ state) would not violate MABK inequalities for $\sin 2\alpha \le 1/\sqrt{2^{N-1}}$. Furthermore,
 in \cite{zuk} authors showed that generalized GHZ states within a specified parameter range for odd number of qubits do
 not violate Werner-Wolf-\.{Z}ukowski-Brukner (WW\.{Z}B) inequalities \cite{brukner}. These inequalities form a complete set of correlation Bell inequalities for $n$ parties, with two measurement settings per party and two outcomes per measurement. So, the question naturally arises whether one can construct some Bell inequalities such that the problematic generalized GHZ state will violate them for the entire parameter range. In this paper, 
 one of our motivations was to construct Bell inequalities to answer this question affirmatively.  This is achieved by making
 different number of measurements on different qubits. It is unlike other previous major inequalities.\\
The second motivation was to attempt to link nonlocality with the entanglement. We  characterize nonlocality of a state
  by the maximum amount
 of violation of a Bell inequality. Both notions of entanglement and nonlocality are fluid for multipartite states. There exist a wide
 array of Bell inequalities, and multiple characterizations of entanglement. In this paper, we are able to link entanglement
 and nonlocality, for the class of generalized GHZ states.
 The third motivation was to be able to discriminate between separable, biseparable, and genuinely entangled pure states using
 Bell inequalities. 
In general, it is very difficult to discriminate between biseparable and genuinely entangled states. MABK inequalities give a sufficient condition to distinguish them \cite{coll,bec,uff}. However as the condition is only sufficient but not a necessary one, biseparable and genuinely entangled states cannot always be distinguished by means of these inequalities. In this work, using our Bell inequalities, one can always distinguish between separable, biseparable and genuinely entangled three-qubit pure states from the pattern of their violations. We have also provided numerical evidence that any pure entangled state will 
violate one or more inequalities from the set. Analytical proof is difficult due to many parameters in the state and
possible measurement settings. Our conjecture is similar in spirit to a few previous works  \cite{bancal}.
Related to this work, in \cite{ch}, some operators have been constructed to distinguish between six SLOCC (Stochastic local operation and classical communication) inequivalent classes for pure three-qubit states.\\ 

The paper is organized as follows. In the next section, we introduce a set of Bell inequalities
for three-qubit states and discuss some of their properties. In the subsequent section, we prove
a number of propositions for three-qubit states. We then generalize these inequalities to the case of $n$ qubits. The last section has conclusions.

\section{A set of Bell inequalities}

We consider a three-qubit system, with a qubit each with Alice, Bob and Charlie. In the
Bell inequalities that we introduce, two of the parties will make two measurements, while
the third party will make only one measurement. This third party can be either Alice, Bob,
or Charlie. A general state need not have any symmetry, therefore we will be considering a
set of Bell inequalities, rather than one inequality. The one measurement by one of the
parties is necessary. We note that in the original Bell inequality \cite{bell1964}, one of the two parties
makes only one measurement.
We will first list the set of six inequalities, and later explain the motivation.
   In this list, the left-hand side should be thought of as the expectation value
   of the observables. In the first and third inequalities, Alice makes one measurement
   given by observable $A_1$, Bob measures the observables $B_1$ and $B_2$,
   and Charlie measures observables $C_1$ and $C_2$. These are dichotomic 
   observables, with values $\{-1, 1\}$.
  In the inequalities $(2)$ and $(6)$,
   Bob measures only one observable,  $B_1$, while in the inequalities $(4)$ and $(5)$,
   Charlie measures only one observable, $C_1$.  Other parties measure two observables.

\begin{eqnarray}
A_1 B_1 (C_1+C_2)+B_2(C_1-C_2) & \leq & 2, \\
A_1 B_1 (C_1+C_2)+A_2(C_1-C_2) & \leq & 2,  \\ \nonumber \\  (B_1+B_2) C_1 + A_1 (B_1-B_2) C_2& \leq & 2, \\
A_1 (B_1+B_2)+A_2(B_1-B_2) C_1& \leq & 2, \\ \nonumber  \\
(A_1+A_2)B_1+(A_1-A_2)B_2C_1 & \leq & 2, \\
(A_1+A_2)C_1+(A_1-A_2)B_1C_2 & \leq & 2.
\end{eqnarray}

To find the maximal violation of these inequalities for a state, one has to consider
  all possible measurements. Therefore, inequalities obtained by interchange of $A_1$ and $A_2$
  will give identical maximal violation. Same will be true for other set of observables. Because of this,
  we do not include such inequalities in our set. In quantum mechanics, the maximal value of these Bell operators
  can be $2\sqrt{2}$. This has been proved along the same line as the Tsirelson's bound for CHSH operator.

\subsection{Quantum bound for the inequalities} 
We will obtain the bound for the first inequality and the analysis will be similar for others.
Let us call the corresponding Bell operator for the first inequality as,
\begin{equation}
B_{3} = A_1 B_1 (C_1+C_2)+B_2(C_1-C_2)
\end{equation}
If we take the square of this expression we get, 
\begin{equation}
B_{3}^{2}= 4I+A_1[C_1,C_2][B_1,B_2].
\end{equation}
Here, we have used ${A_1}^2={B_1}^2={B_2}^2={C_1}^2={C_2}^2=I$.
We know that, for two bounded operators $X$ and $Y$,
\begin{equation}
\parallel [X,Y] \parallel \leq  2 \parallel X \parallel \parallel Y \parallel,
\end{equation}
where, "$\parallel$ $\parallel$" is the sup norm of a bounded operator.
Using this relation, we notice that the maximum value will be obtained when
 $B_{3}^2$  is $8I$ and hence $\parallel B_{3} \parallel \leq 2\sqrt{2}$. This proves our claim. 

\subsection{Motivation behind the inequalities}
To motivate these inequalities, 
our starting point will be the CHSH inequality. This inequality reads as, $A_1B_1+A_1B_2+A_2B_1-A_2B_2\leq 2$, 
where $A_1$, $A_2$ are the measurement observables for Alice, $B_1$, $B_2$ are the measurement observables 
for Bob and $2$ is the local-realistic value. Again in left-hand side, expectation value is implicit.
From Tsirelson's bound \cite{tsirelson}, maximum value of this operator can 
achieve for quantum states is $2\sqrt{2}$. This value is achieved for the maximally entangled
states - Bell states. Let us consider the state $\ket{\phi^+} = \frac{1}{\sqrt{2}}(\ket{00}+\ket{11})$.
This state is useful for generalization to GHZ state. For a suitable choice of measurements, for example, 
$A_1=\sigma_x$, $A_2=\sigma_z$, $B_1=1/\sqrt{2}(\sigma_x+\sigma_z)$ and $B_2=1/\sqrt{2}(\sigma_x-\sigma_z)$, 
we obtain the maximal violation of $2\sqrt{2}$. For this choice of measurements,
the CHSH operator takes the form $\sqrt{2}(\sigma_x\otimes\sigma_x+\sigma_z\otimes\sigma_z)$. 
The state $\ket{\phi^+}$ is its eigenstate with eigenvalue $2\sqrt{2}$ \cite{braunstein}. With local unitary
transformations, we can find other forms of this operator, of which other Bell states will be eigenstates
with maximal eigenvalue. Now, 
we want to construct an operator for three-qubit pure states such that, the GHZ state of three qubits 
will be  the eigenstate of this operator with highest eigenvalue. Like the CHSH operator, we can 
construct an operator such that its maximum eigenvalue will be $2\sqrt{2}$. The GHZ state, 
$\frac{1}{\sqrt{2}}(\ket{000}+\ket{111})$, is the eigenstate of  the operator 
$\sqrt{2}(\sigma_x \otimes \sigma_x \otimes \sigma_x + \sigma_z \otimes \sigma_z \otimes I)$
with eigenvalue $2\sqrt{2}$. We can write other forms of this operator where identity operator 
acts on other qubits.
 We clearly see that we have even number of $\sigma_z$; here it
is one fewer than the number of qubits. This suggests that we need to make only one
measurement on one of the qubits. With the help of this operator, we can construct
the simplest set of Bell inequalities. We need a set to take care of asymmetric situations.
This set is given above. To look at it from a different point of view,
identity in one place of the aforementioned operator gives us hint to construct non-correlation 
Bell inequality. Also from previous discussion, it is clear that to obtain violations for 
all pure entangled states, correlation Bell inequalities are not enough. So, it seems that 
non-correlation Bell inequalities may work. We show below by first considering generalized
GHZ states and then arbitrary three-qubit states that it is indeed true.
\section{Three-qubit states}

    In this section, we will consider three different classes of states -- product states, pure biseparable
    states, and states with genuine tripartite entanglement. We will see how our inequalities can
    distinguish these classes of states. In addition, we shall consider generalized GHZ states. Theses states 
    are symmetric under the permutation of particles; so we can pick any of the inequalities. All
    will be violated in the same manner. \\
    
  {\em \bf Proposition 1:}
 {\it All  generalized GHZ states violate all six inequalities of this set}.\\
 
\textbf{Proof:} Let's consider the three-qubit generalized GHZ state,
\begin{equation}
\label{GGHZ}
\ket{GGHZ}=\alpha \ket{000}+\beta \ket{111}.
\end{equation}

\begin{figure}
\begin{center}
\includegraphics[height=3.2cm,width=6.4cm]{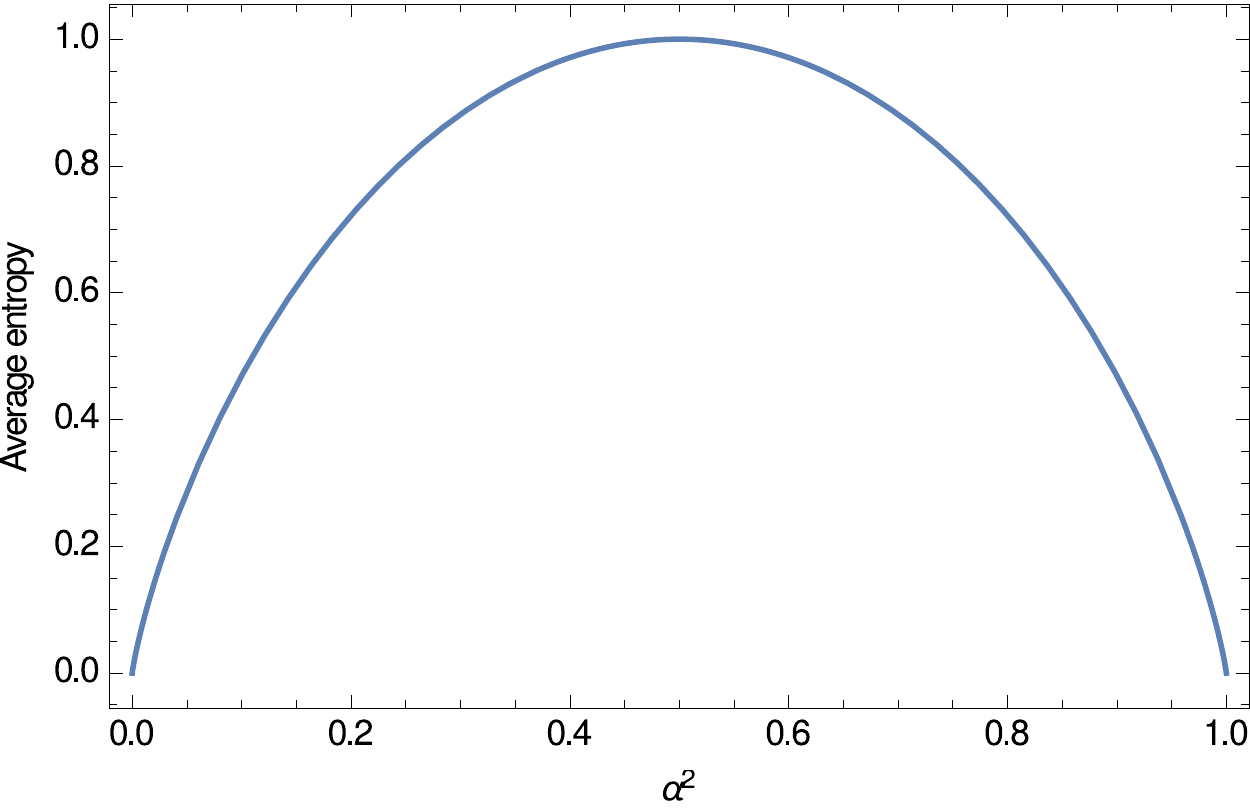}
\caption{ Average Von Neumann entropy over the three bipartitions vs $\alpha^2$ plot.}
\label{en}
\end{center}
\end{figure}

     These states have been problematic for different inequalities. However, as our Bell inequalities
     were designed for GHZ states, all of these generalized GHZ states violate all our inequalities.
     In the spirit of generalized Schmidt decomposition, we can take $\alpha$ and $\beta$ to
     be real and positive numbers. Quantification of entanglement in multipartite scenario is a messy business. Unlike pure bipartite system, there is no unique measure of entanglement for multipartite states \cite{pleino, multi}. One uses different measures depending upon different purposes. Von Neumann entropy uniquely captures and quantify the entanglement for a pure bipartite system in the asymptotic limit. For a pure multipartite state one can use the average of Von Neumann entropy over each bipartition as a suitable measure of multipartite entanglement \cite{multi}.   For three qubit pure state, there are three bipartitions namely, $1-23$, $2-31$ and $3-12$. Average of Von Neumann entropy for generalized GHZ state as defined in equation (\ref{GGHZ}) over these bipartitions is $-\alpha^2 \log_2 {\alpha^2}-\beta^2 \log_2 {\beta^2}$. This is also the entropy for each bipartition for these states as the states are symmetric. We have plotted this average entropy with $\alpha^2$ with $\alpha^2+\beta^2=1$, in figure (\ref{en}).

\begin{figure}
\begin{center}
\includegraphics[height=3.2cm,width=6.4cm]{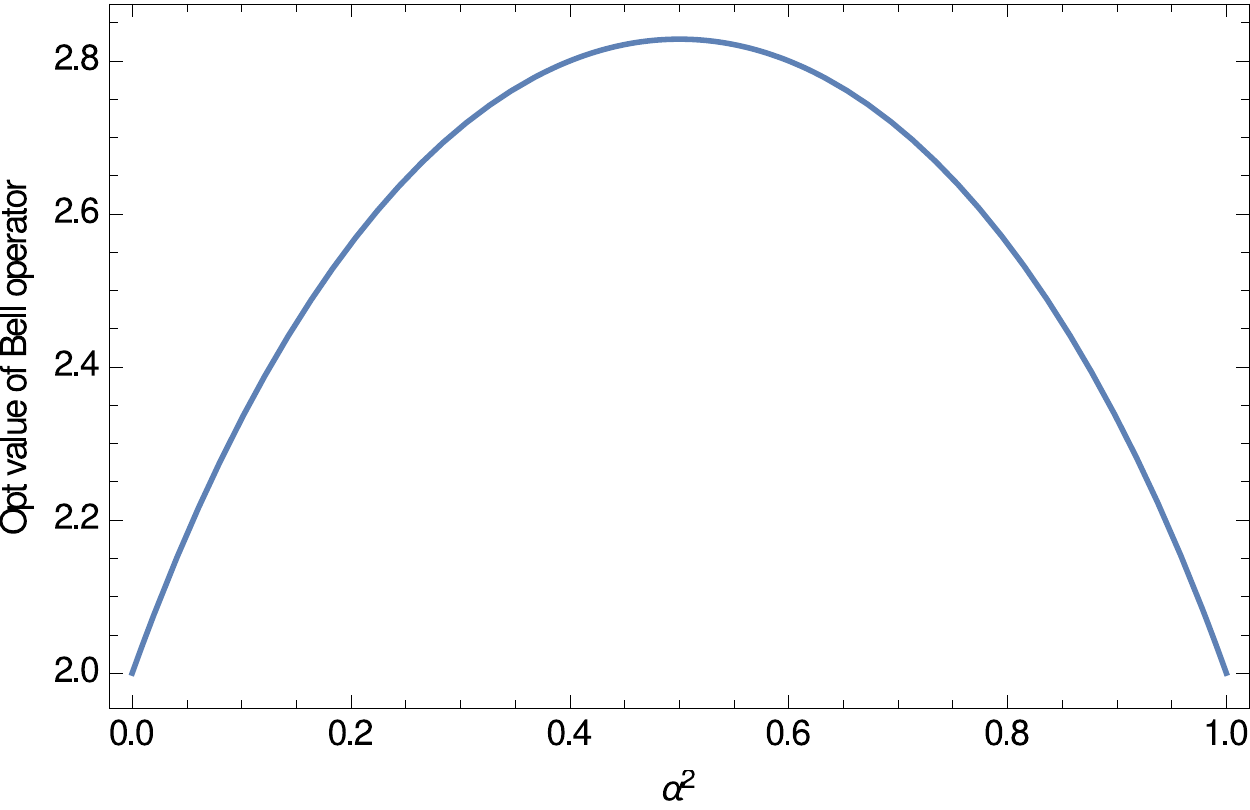}
\caption{ Maximum expectation value of the Bell operator for a generalized GHZ state vs $\alpha^2$ plot.}
\label{non}
\end{center}
\end{figure}

     Since the states are symmetric under the permutations of particles, we can choose any Bell
     inequality from the set. We choose  the inequality,
\begin{equation}
A_1 (B_1+B_2)+A_2(B_1-B_2)C_1\leq 2.
\end{equation}
Let us recall that the expectation value for the left-hand side is implicit.
We choose the following measurement settings,
$A_1=\sigma_z$, $A_2=\sigma_x$, $B_1=\cos\theta\sigma_x+\sin\theta\sigma_z$, $B_2=-\cos\theta\sigma_x+\sin\theta\sigma_z$, $C_1=\sigma_x$.
For these measurement settings, the expectation value of the above Bell operator for the
generalized GHZ state is 
\begin{equation}
\bra{GGHZ}A_1 (B_1+B_2)+A_2(B_1-B_2)C_1\ket{GGHZ}.
\end{equation}
Its value is $2[2\alpha\beta\cos\theta+(\alpha^2+\beta^2)\sin\theta]=2[2\alpha\beta\cos\theta+\sin\theta]$. 
Now, $a\sin\phi+b\cos\phi\leq\sqrt{a^2+b^2}$.
Therefore
$\bra{GGHZ}A_1 (B_1+B_2)+A_2(B_1-B_2)C_1\ket{GGHZ}$ is less than or equal to $2\sqrt{1+4\alpha^2\beta^2}$,
which is always greater than $2$ for nonzero $\alpha$, $\beta$ and gives maximum value $2\sqrt{2}$ 
for the conventional GHZ state. The upper bound on the expectation value can be written as 
$2\sqrt{1+ {\cal C}^2}$, where ${\cal C}^2 = 4 {\alpha}^2 {\beta}^2$ is nothing but the tangle of the generalized GHZ state.
The quantity
${\cal C}$ is also like concurrence for a two-qubit bipartite state. We have also plotted the optimized expectation value of the Bell operator with $\alpha^2$ in figure (\ref{non}).
From these two plots, it is clear that, the entanglement measure (average Von Neumann entropy over the bipartitions) and the maximum amount of Bell violation for generalized GHZ states are monotonically related to each other. 
In a different way, we can say that for the generalized GHZ state, the expectation value of the Bell operator depends on the amount of entanglement. The more is the entanglement of a state, the more Bell nonlocal it is. This concludes the proof.

As discussed earlier, this is the class of states which was creating problem for MABK, Svetlichny and moreover for all 
correlation Bell inequalities for a particular range of $\alpha$ and $\beta$. However,
all states in this class violate all the inequalities in our set. \\

  {\em \bf Proposition 2:}
 {\it Any separable pure three-qubit state obeys all the inequalities within the set.} \\
 
\textbf{Proof:} Our
  inequalities are the hybrid version of the CHSH inequality. All separable pure three-qubit 
states can be written, after applying some convenient local unitary transformation as $\ket{0}\ket{0}\ket{0}$.
 We can use any one of the 
operators, say $A_1B_2(C_1+C_2)+B_1(C_1-C_2)$, as proof will be same for all the others. We take 
most general quantum measurements for the five observables,
 as $A_1=\sin{\theta_1}\cos{\phi_1}\sigma_x+\sin{\theta_1}\sin{\phi_1}\sigma_y+\cos{\theta_1}\sigma_z$ 
and similarly for other observables $A_2$, $B_1$, $B_2$, $C_1$, $C_2$, for which the parameters are 
$\theta_2$, $\phi_2$; $\theta_3$, $\phi_3$; $\theta_4$, $\phi_4$; $\theta_5$, $\phi_5$; $\theta_6$, $\phi_6$ 
respectively. Now, we calculate the expectation value of the operator for the separable state. 
Putting the form of the measurement for each operator, we get the expectation value to be 
$\cos\theta_2(\cos\theta_5-\cos\theta_6)+\cos\theta_1\cos\theta_3(\cos\theta_5+\cos\theta_6)$. 
Obviously we can write this expression as the conventional CHSH operator like $X_1(Y_1+Y_2)+X_2(Y_1-Y_2)$, 
where $X_1=\cos\theta_1\cos\theta_3$, $X_2=\cos\theta_2$, $Y_1=\cos\theta_5$, $Y_2=\cos\theta_6$,
and $X_1, X_2, Y_1, Y_2\leq 1$. So, clearly there will be some values of these operators such that 
the operator takes the maximum possible value, which is $2$ (in analogy with the local-realistic value 
of CHSH operator). Similarly, for all other inequalities in the set, we will get the optimum violation of
$2$ for all separable states. This concludes the proof. \\

{\em \bf Proposition 3:}
{\it All biseparable pure three-qubit states violate exactly two inequalities within the set and the amount of maximal violation are same for both.}\\

\textbf{Proof:} This can be proved by observing the form of the inequalities. 
We can rewrite any biseparable state as an equivalent form of $\ket{0}(\alpha\ket{0}\ket{0}+\beta\ket{1}\ket{1})$ by local unitary transformations. (We can relabel qubits such that number `2' and `3'
are entangled.
This state is separable in $1-23$ bipartition. So, those inequalities, which can explore the entanglement 
between the second and the third qubit will be violated. For example, for the above mentioned state, 
$A_1B_1(C_1+C_2)+B_2(C_1-C_2)\leq 2$ will be violated, because  a CHSH type operator for second 
and third qubits is embedded in this operator. So, the amount of violation will be exactly 
same as in the case of two-qubit entangled state  
and the CHSH operator. Not only this inequality, but there is another inequality within this set, which will also be violated in this case. This inequality is,
$(B_1+B_2)C_2+A_1(B_1-B_2)C_1  \leq  2.$
So, there are two inequalities, which 
will be violated for a given pure biseparable state. Also, as all the two operators have the same 
form (the CHSH form) in second and third particle, the amount of maximal violations will be same in 
two cases. And the last important fact is that, no other states (except biseparable pure states) 
will have same kind of violations, i.e exactly two violations of the same maximal amount. This concludes the proof.\\

Until now, we have considered special classes of three-qubit states. One would like to show that
any genuinely entangled tripartite state will violate one of our inequalities. For this, we will be
presenting numerical evidence, using a general parametrized form of a three-qubit state.

{\em \bf  Proposition 4:}
{\it For all genuine tripartite entangled states, we have violation within the set.}

\begin{figure}[H]
\begin{center}
\includegraphics[height=2.9cm,width=6.4cm]{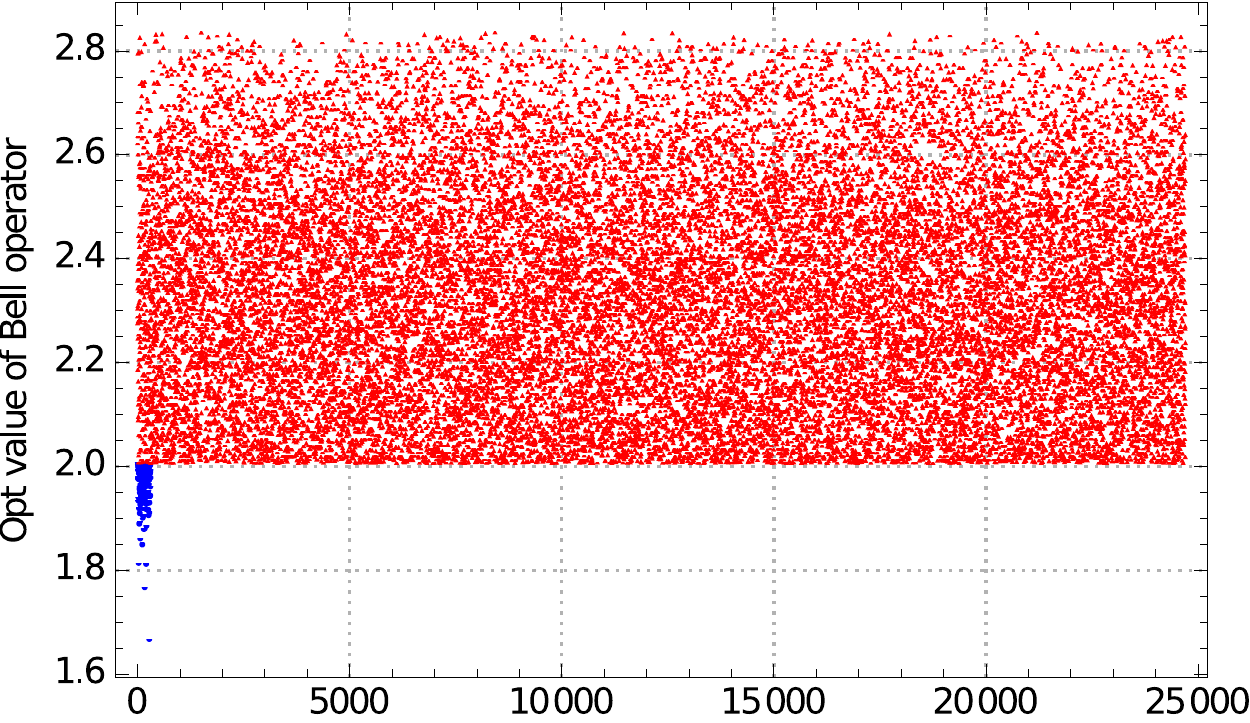}
\caption{Optimum value of the Bell operator $(1)$. Out of 25000 states, 297 states do not violate this inequality. States which violate the inequality are shown by red points and those do not are shown by blue points.}
\label{first}
\end{center}
\end{figure}

\begin{figure}[H]
\begin{center}
\includegraphics[height=3.0cm,width=6.4cm]{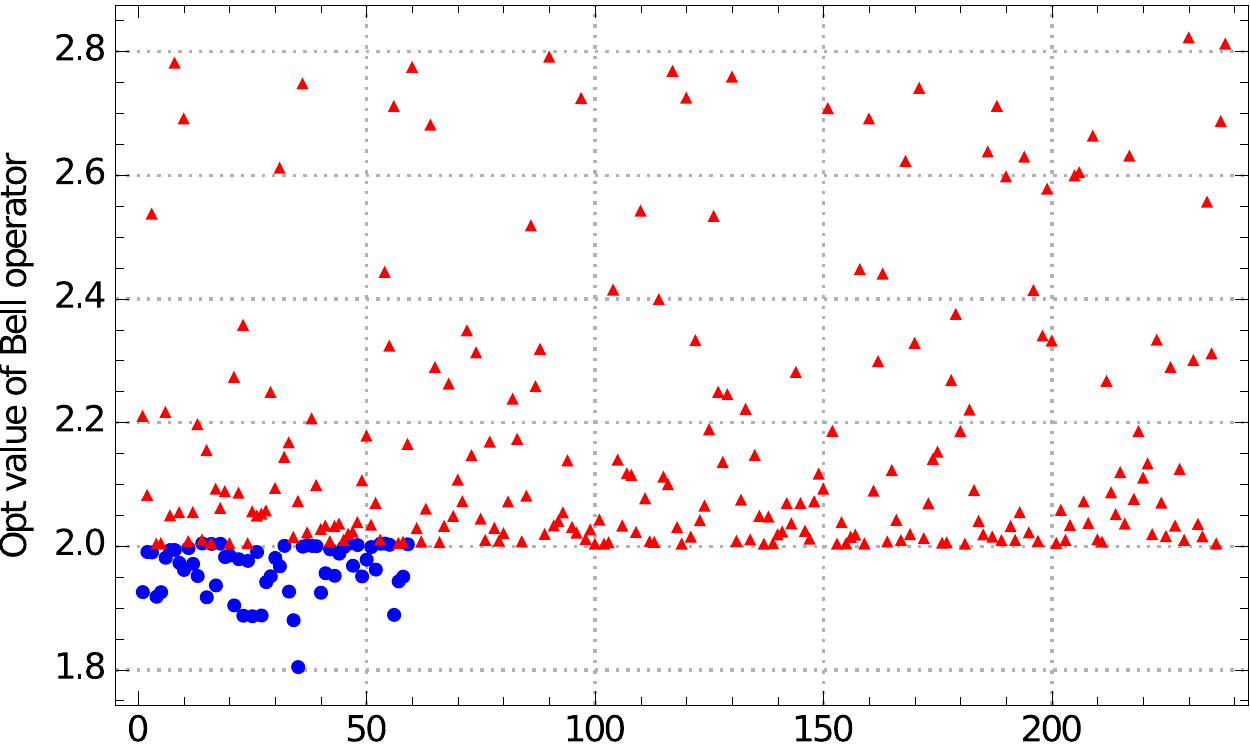}
\caption{Optimum value of the Bell operator $(2)$. Out of 297 states, 59 states do not violate this inequality. States which violate the inequality are shown by red points and those do not are shown by blue points.}
\label{ineq2}
\end{center}
\end{figure}

\begin{figure}[H]
\begin{center}
\includegraphics[height=3.0cm,width=6.4cm]{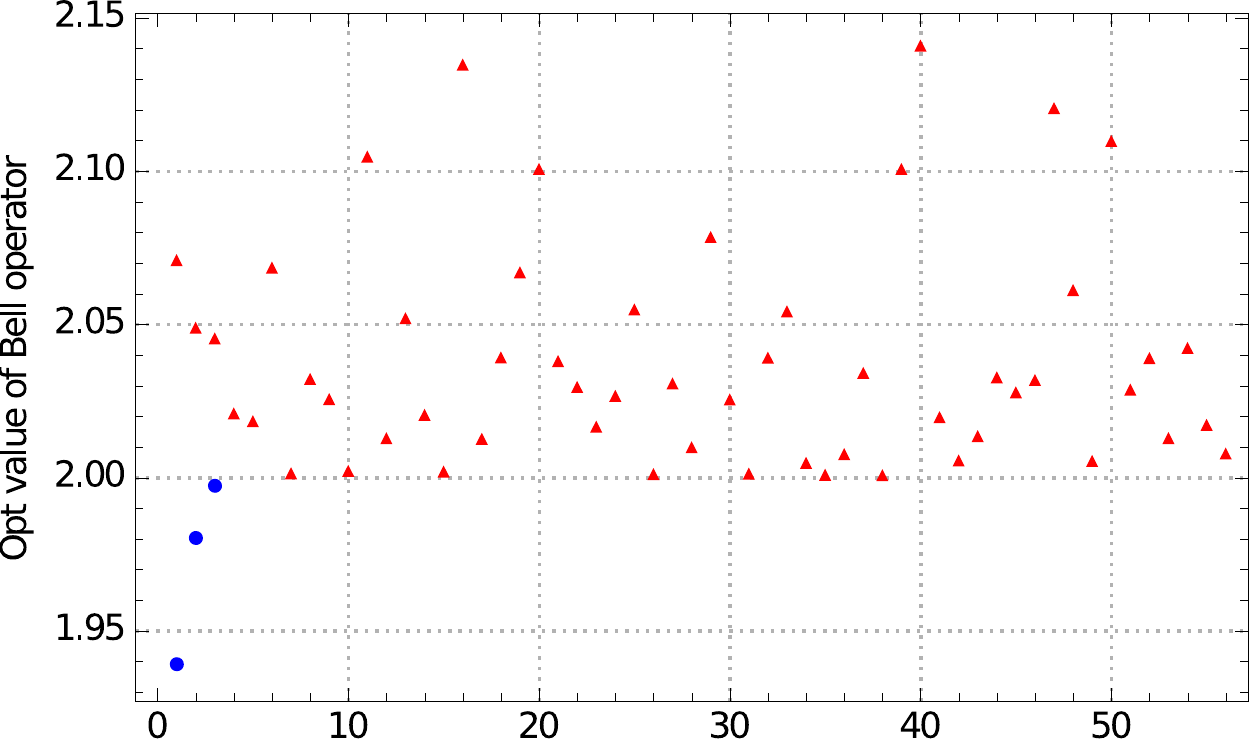}
\caption{Optimum value of the Bell operator $(3)$. Out of 59 states, 3 states do not violate this inequality. States which violate the inequality are shown by red points and those do not are shown by blue points.}
\label{third}
\end{center}
\end{figure}

\begin{figure}[H]
\begin{center}
\includegraphics[height=3.0cm,width=6.4cm]{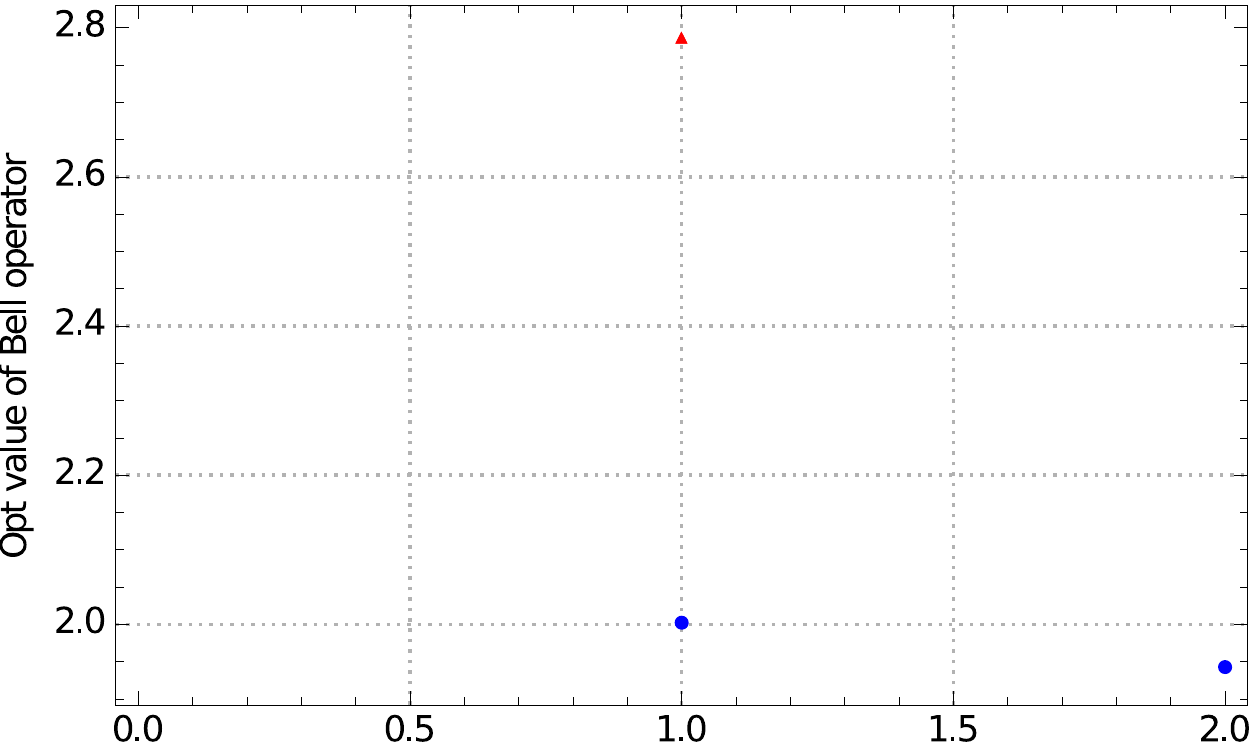}
\caption{Optimum value of the Bell operator $(4)$. Out of 3 states, 2 states do not violate this inequality. States which violate the inequality are shown by red points and those do not are shown by blue points.}
\label{fourth}
\end{center}
\end{figure}

\begin{figure}[H]
\begin{center}
\includegraphics[height=3.0cm,width=6.4cm]{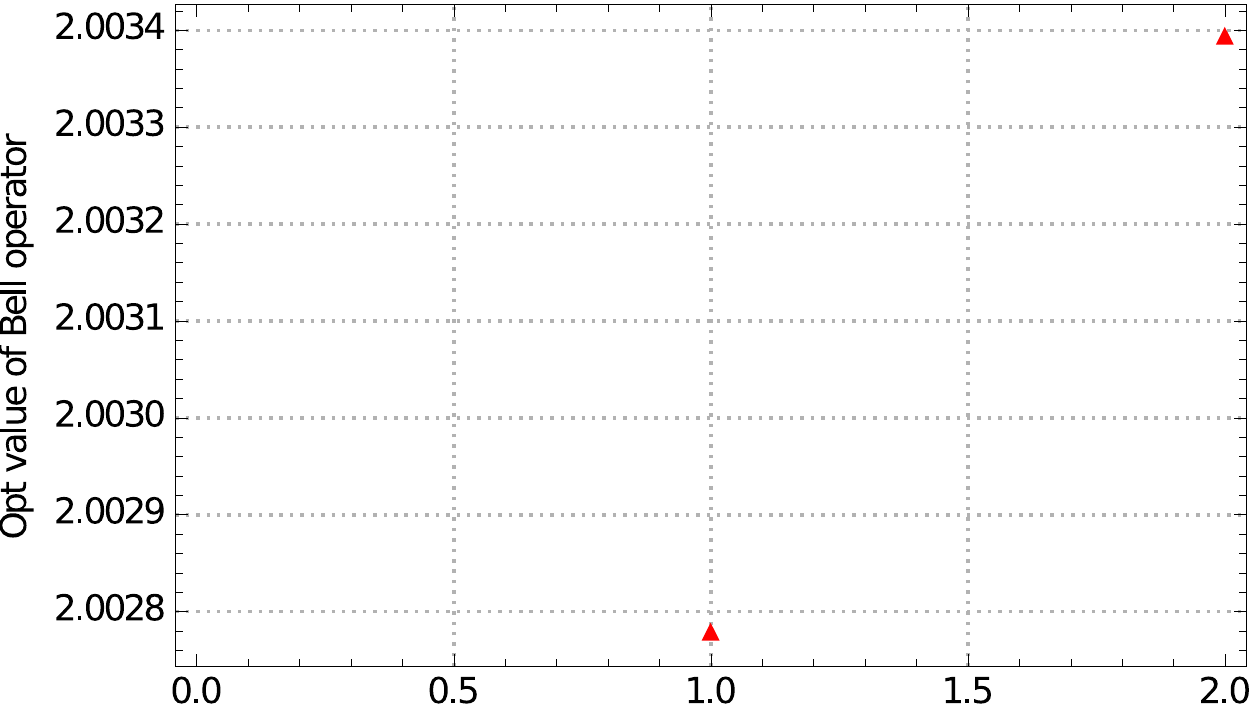}
\caption{Optimum value of the Bell operator $(5)$. Out of 2 states, all the states violate this inequality. So, there are 2 red points and no blue points.}
\label{fifth}
\end{center}
\end{figure}

We do not have an analytical proof for this proposition. But we present supporting numerical evidence. 
Any genuinely entangled  three-qubit pure state can be written in a canonical form \cite{14} with six parameters. 
This form includes the GHZ and W class states \cite{dur} for three qubits. For biseparable pure states, 
we have already provided proof for the violation of inequalities within the set.  The canonical form of a
general three-qubit state is,
\begin{multline}
\ket{\psi}=\lambda_0\ket{0}\ket{0}\ket{0}+\lambda_1 e^{i\phi}\ket{1}\ket{0}\ket{0}+
\lambda_2\ket{1}\ket{0}\ket{1}\\
+\lambda_3\ket{1}\ket{1}\ket{0}+\lambda_4\ket{1}\ket{1}\ket{1}, 
\end{multline}
where $\lambda_i\geq 0$, $\sum_i{\lambda_i}^2=1$, $\lambda_0\neq 0$, $\lambda_2+\lambda_4\neq 0, 
\lambda_3+\lambda_4\neq 0$ and $\phi\in [0,\pi]$.
We have randomly generated 25,000 states and tested our set of Bell inequalities. The expectation value of a
 Bell operator is optimized by considering all possible measurement settings for all observables. Starting from the inequality $(1)$ from the set, we continued with other inequalities one after one until all the generated states violate one inequality from the set. Results are displayed in figures (\ref{first}-\ref{fifth}).
At first, Bell inequality $(1)$ was tested for these randomly generated states and out of 25000 states, 297 states do not violate this inequality, as shown in figure (\ref{first}).
Then using the inequality $(2)$ with these 297 states, number of states which do not violate these first two inequalities was further reduced to 59 states, as shown in figure (\ref{ineq2}).
Similarly, applying the other Bell inequalities from the proposed set one by one the number of states showing no violation for those inequalities can be reduced to zero. 
We have shown in figures (\ref{first}-\ref{fifth}), starting from 25000 random states, violation for each state has been obtained using first five inequalities from the proposed set.

However, this random generation of states would not be setting any of the parameters as zero. But we should consider those states also for numerical checking. So, we have generated 5000 states each for 9 more classes of states, by setting some of the parameters as zero.
These classes are obtained as, only $\lambda_1=0$, only $\lambda_2=0$, only $\lambda_3=0$, only $\lambda_4=0$, only $\lambda_1, \lambda_2=0$, only $\lambda_1, \lambda_3=0$, only, $\lambda_1, \lambda_4=0$, only, $\lambda_2, \lambda_3=0$, Only, $\lambda_1, \lambda_2, \lambda_3=0$. For each of theses class, $\phi$ is arbitrary. We have taken $5000$ random values of each parameter for each class and found violations within the set of $6$ inequalities in each case. Based on this and the fact that all generalized GHZ class states violate each inequality (already proved), we  expect that this set can certify genuine pure tripartite entanglement.

To conclude the case of tripartite scenario, we have established that all generalized GHZ states violate all the inequalities within the set and with the help of propositions 1,2 and 3 one can always distinguish between separable, biseparable and genuinely entangled pure states from the pattern of their violations of inequalities from the set.
\section{Multi-qubit states}

    We have established that our set of inequalities are violated by any entangled three-qubit
    state. We can generalize this set of inequalities to $n$-qubit states. 
This extension  for multi-qubit scenario is straight-forward. One will have to distinguish
between two cases --  odd number of qubits and even number of qubits. Starting from the 
operator, of which GHZ state is an eigenstate, one can construct different Bell inequalities. For even $n$, there will be
a set of $n$ inequalities; while for odd $n$, the number will rise to $ n(n-1)$. The set is larger
for odd number of qubits, because we have choice of making one measurement on any of
$n$ qubits; while in the case of even $n$, two measurements are made on all qubits.
Therefore, we have to construct different types of inequalities for even and odd number of particles. 
We have already seen that the GHZ state of three qubits is eigenstate of the operator 
$\sqrt{2}(\sigma_x \otimes \sigma_x \otimes \sigma_x + 
\sigma_z \otimes \sigma_z \otimes I)$ with 
the highest eigenvalue $2\sqrt{2}$. This form of the operator can be generalized for any 
$n$-qubit GHZ state, when $n$ is odd. $n$-qubit GHZ states is the eigenstate of the operator
$\sqrt{2}(\sigma_x \otimes \sigma_x \otimes \sigma_x\otimes\cdots\otimes\sigma_x^{nth} + 
\sigma_z \otimes \sigma_z \otimes\cdots\otimes\sigma_z^{(n-1)th}\otimes I)$ 
with the highest eigenvalue $2\sqrt{2}$. So, like the three-qubit
case, we have to consider non-correlation Bell inequalities when $n$ is odd. The
first two Bell inequalities $(1)$ and $(2)$ can be easily generalized  
 for $n$-qubit pure states as,

\begin{multline}
 A_1 A_2  A_3 A_4A_5.. (A_{n}  + {A'_n}) +\\
{A'_2}  {A'_3} {A'_4} {A'_5}..(A_{n} -{A'_n})  \leq   2, 
\label{gen1}
\end{multline}
and
\begin{multline}  
A_2  A_3 A_4A_5.. (A_{n}  + {A'_n}) +\\
A_1{A'_2}  {A'_3}{A'_4}{A'_5}..(A_{n} -{A'_n})  \leq 2.  
\label{gen2}
\end{multline}

Here, $A_i$ and ${A'_i}$ are two dichotomic observable for $i^{th}$ party.  In these
inequalities, one measurement has been made on first qubit. Similarly one can make
single measurement on $(n-2)$ other qubits. This will lead to $(n-1)$ inequalities.
We can write $n$ such $(n-1)$ inequalities with $(A_{i}  \pm {A'_i})$ for $i^{th}$
qubit, giving a set of total $n(n-1)$ inequalities. For three-qubits the number of inequalities in the set is six. For finding maximal violation,
we consider all allowed $A_i$ and ${A'_i}$, therefore their positions in the 
inequalities can be interchanged. The above set of inequalities can be used
to characterize the entanglement of $n$-qubit states for odd $n$. In the case of generalized $n$-qubit GHZ states, any one of these generalized inequalities is 
enough. One can show that for odd number of qubits these non-correlation Bell inequalities are 
violated by all generalized GHZ states with maximum violation of $2\sqrt{2}$ for the conventional 
GHZ state. The proof is similar to the three-qubit case. 
Situation changes when one considers GHZ like states with even number of qubits. Because now, 
like the Bell states, the conventional GHZ state of $n$ qubits ($n$ is even) is the 
eigenstate of the operator $\sqrt{2}(\sigma_x \otimes \sigma_x \otimes \sigma_x\otimes\cdots\otimes\sigma_x^{nth} + 
\sigma_z \otimes \sigma_z \otimes\cdots\otimes\sigma_z^{(n-1)th}\otimes \sigma_z)$ with highest 
eigenvalue $2\sqrt{2}$. This suggests that correlation Bell inequalities are required in this case. 
For example, one can generalize the first correlation Bell inequality as,
\begin{multline}
(A_1+{A'_1}) A_2 A_3 A_4A_5..A_n+\\
(A_1- {A'_1}) {A'_2}{A'_3}{A'_4}{A'_5}..{A'_n}\leq 2.
\label{gen3}
\end{multline}
Similarly, $n$ such inequalities with $(A_{i}  \pm {A'_i})$ can be written.
Again, among these correlation Bell inequalities any one of them can be used for generalized GHZ states.
 The proof that any generalized GHZ state with even number of qubits violate these inequalities 
 can be carried along the same line as for the 
three-qubit case. The fact that generalized GHZ states with even number of qubits violate a 
correlation Bell inequality within the set of all correlation Bell inequalities \cite{brukner} was 
known \cite{zuk}. But it is important to note that, the correlation Bell inequality violated by the generalized GHZ state with even number of qubits, may not be MABK inequalities. Here, we have introduced a set of correlation Bell inequality 
which must be violated by all generalized GHZ states with even 
number of qubits. Like three-qubit states, one may expect that any $n$-qubit pure state for
odd value of $n$ will violate one of the $ n(n-1)$ inequalities  like in $(\ref{gen1})$ and $(\ref{gen2})$, 
while for even $n$,
one of the $n$ inequalities like in $(\ref{gen3})$ will be violated. \\
   
  {\em \bf Proposition 5:}
 {\it Multiqubit extension of the inequalities are violated by multiqubit generalized GHZ states.}\\

\textbf{Proof:}
 Let's consider the generalized $n$-qubit GHZ state
\begin{equation}
\ket{GGHZ}_{n} = \alpha \ket{00.....00} + \beta \ket{11.....11}.
\end{equation}
  In this state, first term represents all $n$ qubits in the `0' state and the second term
  is for all $n$ qubits in the `1' state. The proof will follow exactly same steps as in the
  proposition 1. The results will also be identical. In the case of both even and odd $n$,
  the maximal violation would be $2\sqrt{1+ {\cal C}^2}$, where ${\cal C} = 2 \alpha \beta$.
  For the $n$-qubit GHZ state ${\cal C} = 1$ and the maximal violation will be
  $2\sqrt{2}$.
  
\section{Conclusion}  

We have presented a new set of six  Bell inequalities. Separable 
three-qubit pure states do not violate any of these inequalities and biseparable pure three-qubit 
states violate exactly two of them with same maximal amount. A generalized GHZ state violates 
all the inequalities in the set, with conventional GHZ state giving maximum amount of violation, 
which is $2\sqrt{2}$.  Furthermore, for this class of states, our inequalities provide a
link between nonlocality and entanglement. More entangled state will violate the inequalities more.
We have also provided numerical evidence that any genuine tripartite 
entangled pure state will give violation within this set. A key point of this set of inequalities
is that one will make only one measurement on one of the qubits. For violation this
measurement is necessary. It is similar to the original Bell inequality. 
It can also be used to distinguish between separable, biseparable and genuinely entangled pure three qubit states. One can also examine the
three-qubit mixed states, where one may expect to find the phenomenon of hidden nonlocality with respect to our set of inequalities.
These inequalities have also been generalized for multi-qubit 
scenario. Each of these inequalities will be violated by a  generalized multi-qubit GHZ state.
It is highly likely that a set of inequalities similar to three-qubit states can detect and
characterize the entanglement of multi-qubit states. However in the absence of a
parametrized form of a pure entangled states beyond three-qubit case, 
we cannot do numerical analysis for the whole set.


\end{document}